# Einstein's Oxford Blackboard: A unique historical artefact

Cormac O'Raifeartaigh

Albert Einstein's blackboard is a well-known exhibit at the History of Science Museum in Oxford. However, it is much less well known that the writing on the board provides a neat summary of a work of historic importance, namely Einstein's 1931 model of the expanding Universe. As a visual representation of one of the earliest models of the Universe to be proposed in the wake of Edwin Hubble's observations of the recession of the nebulae, the blackboard provides an intriguing snapshot of a key moment in modern cosmology. In addition, one line on the blackboard that is not in Einstein's 1931 paper helps explain some anomalies in the estimates of the size and density of the Universe found in that paper.

## 1. Introduction

The blackboard used by Albert Einstein (1879–1955) on 1931 May 16 in the second of three Rhodes Memorial Lectures at Oxford University is a well-known historical artefact that has been on display at the History of Science Museum, Oxford, for many years.[1] Indeed, it is one of the museum's most famous exhibits.[2] The blackboard's provenance is well known and has been described in the literature.

However, surprisingly little attention has been paid to the science underpinning the writing on the blackboard. For many years, the caption attached to the exhibit read 'based on an unknown work by Albert Einstein'. Similarly, a short BBC film describing the blackboard makes no reference to the physics underlying the work.[3]

A few years ago, this author pointed out that the writing on the board is taken directly from Einstein's paper *Zum kosmologischen Problem der allgemeinen Relativitätstheorie*, submitted for publication in 1931 April.[4,5] This paper is of considerable historical importance as it constitutes Einstein's first published model of the expanding Universe. Thus, the blackboard provides an intriguing snapshot of one of the great paradigm shifts of 20th-century science, the period during which theorists such as Einstein transitioned from static to expanding cosmologies in the wake of the emergence of the first astronomical evidence for cosmic expansion.

In addition, the blackboard is not a passive historical artefact. Although most of the writing on the board repeats the analysis of Einstein's 1931 paper, there is one line on the board that is not in the paper. As we shall see later, this line gives an important insight into a puzzling systematic error in Einstein's use of astronomical data to extract numerical estimates for his cosmic model. Thus, Einstein's Oxford blackboard is both a museum exhibit of significant historical interest and an active historical artefact that contributes to our understanding of a pivotal moment in the history of science.

## 2. Historical context

In 1931 May, Einstein spent over three weeks at Oxford University at the invitation of the Oxford physicist Frederick Alexander Lindemann (1886–1957). The main purpose of the visit was for Einstein to give a series of three prestigious lectures known as the Rhodes Memorial Lectures, and to be conferred with an honorary degree of Doctorate of Science by the university.

The three Rhodes lectures were duly given on 1931 May 9, 16, and 23. Many historians have noted that the lectures were not an unqualified success, as they were quite technical and delivered in German. While the first lecture attracted a large audience of about 500, the audience dwindled significantly for the remaining two.[6]

As regards the scientific content of the lectures, the first comprised a brief overview of the general theory of relativity, the second presented an overview of relativistic cosmology, and the third gave an overview of Einstein's attempts at a unified field theory.[7] Einstein made copious use of the blackboard during each of his lectures; luckily, one of the blackboards used during the second lecture was preserved by a member of staff.[8] It is this blackboard that is displayed in the History of Science Museum at Oxford (Figure 1 and front cover).

While the first lecture attracted the largest audience,

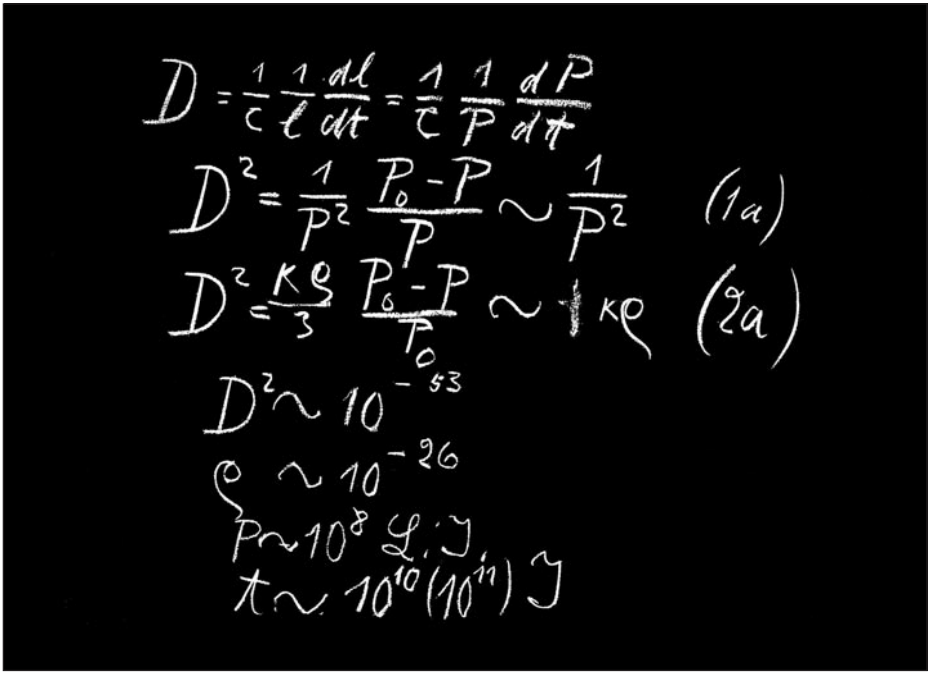

Figure 1: An image of the blackboard used by Einstein in his second Rhodes lecture at Oxford University in 1931 May. The writing on the board presents a brief summary of Einstein's cosmic model of 1931. The fourth line, $D^2 \sim 10^{-53}$, gives valuable insight into a systematic error in Einstein's estimates of cosmic parameters such as the density of matter (the figure rho in the fifth line) and the radius of the Universe (P in the sixth line). In the final line, t is the approximate age of the Universe.
(© History of Science Museum, Oxford).

one suspects the second lecture was of more contemporary interest to physicists in the audience. Following the publication of Edwin Hubble's observations of a linear relation between redshift and distance for the spiral nebulae, many physicists had become interested in the concept of an expanding Universe. In this lecture, Einstein gave a brief summary of a paper on cosmic expansion he had recently submitted for publication.[9]

In the submitted paper, Einstein adopted a dynamic, relativistic model of the Universe first proposed by the Russian mathematician Alexander Friedmann (1888–1925), but simplified the analysis by removing the cosmological constant term from the field equations of relativity, declaring it no longer necessary.[10] This simplification allowed Einstein to extract rough estimates of cosmic parameters such as the density of matter, the radius of the Universe, and the timespan of expansion with the use of Hubble's data. Although the 1931 paper was later eclipsed by a model that was simpler still,[11] the work is of historical importance as Einstein's first model of the expanding Universe (and is sometimes known as the Friedmann–Einstein model).

### 3. The science of the blackboard

Looking at the blackboard, we note that the first equation displayed is

The meaning of this equation is easily discerned from Einstein's 1931 paper.[12] Here, Einstein assumes that Hubble's observations of a linear relation between the distances (l) of the nebulae and their redshifts (interpreted as recession velocities $dl/dt$) is a manifestation of a general expansion on cosmic scales.

To obtain an empirical estimate of the rate of cosmic expansion, he equates the quantity

$D = (1/c).(1/l).(dl/dt)$

with the quantity

$(1/c).(1/P).(dP/dt)$,

where P represents the cosmic radius and $dP/dt$ represents the rate of increase of the radius (in modern parlance, the parameter D is simply the Hubble constant divided by the speed of light or $D = H/c$).

In the next two equations shown on the board, numbered 1a and 2a, Einstein writes

These equations represent expressions for the expansion parameter D derived from theory.

In his 1931 paper, Einstein derived two differential equations for cosmic expansion from the field equations of general relativity in the same manner as Friedmann and the Belgian Georges Lemaître (1894−1966). However, dispensing with the cosmological constant term allowed him to simplify the differential equations considerably and he then expressed the equations in terms of the expansion parameter D as above.

We note that the quantity $P_0$ represents a maximum value for cosmic radius; in his 1931 model, Einstein assumed the Universe to have a closed geometry that would expand to a maximum value and then contract. In both theoretical equations above, Einstein simplifies the analysis even further by setting the quantities $(P_0-P)/P$ and $(P_0-P)/P_0$ to unity.

In the 1931 paper, his justification for this is the assumption that the current cosmic radius is of the same order of magnitude as the maximum value; this approximation confirms the impression that Einstein does not expect accurate results from his analysis, but is merely attempting to obtain rough order-of-magnitude estimates for the parameters of his model.

All that remains is for Einstein to use Hubble's observations to calculate a value for the expansion parameter $D$. Thus the fourth blackboard equation

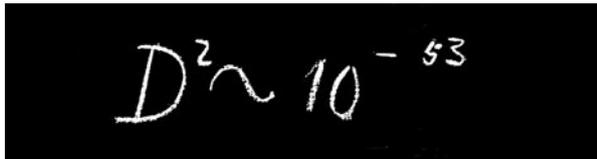

is obtained by substituting Hubble's value for the velocity/distance ratio of the nebulae into the first equation on the board.[13] Assuming this calculation is correct (see Section 4), it is easy to see how substituting this value for $D^2$ into the third equation on the board gives the estimate for the density of matter, rho ($\rho$), of

$\rho \sim D^2/\kappa \sim 10^{-26}$

as shown on the board (here we have taken the Einstein constant as $\kappa = 8\pi G/c^2$ or $1.866 \times 10^{-27}$ cm/g).

Similarly, it is easy to see how Einstein substitutes his empirical value for $D^2$ into the second equation on the board to obtain the order-of-magnitude estimate

$P \sim \sqrt{1/D^2} \sim 10^8$ LJ

for the present radius of the cosmos, expressed in light years (the initials LJ on the board stand for 'Lichtjahr', German for light year). The analysis finishes with an estimate of the time of expansion, $t$, in years although it is not entirely clear how Einstein arrives at this figure (see below). As these are all rough estimates, the second figure in brackets presumably means 'at most $10^{11}$ years'.

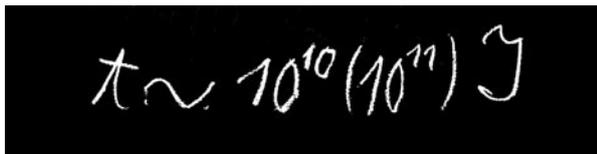

## 4. The blackboard solves a riddle

The outstanding aspect of Einstein's 1931 model of the Universe is the removal of the cosmological constant term from his cosmology. Quite apart from aesthetic and philosophical considerations, this step allowed him to use Hubble's data to obtain rough estimates of cosmic parameters such as the density of matter, the present radius of the Universe, and the age of the Universe.

However, while the calculations on the board (and in the 1931 paper) appear to be self-consistent, they are not consistent with estimates of the same parameters from contemporaneous cosmic models. For example, in the famous Einstein–de Sitter model of 1932,[14] the present radius of the Universe is estimated to be about $2 \times 10^{27}$ cm (or $2 \times 10^9$ light years), a figure that is an order of magnitude larger than that of the blackboard (and the 1931 paper). Similarly, the density of matter is estimated at $1 \times 10^{-28}$ cm$^{-3}$, a figure that is two orders of magnitude smaller than that of the blackboard (and the 1931 paper).

Unfortunately, it is quite hard to pinpoint the source of the error in the paper, as the section where Einstein uses astronomical observations to extract values for cosmic parameters is rather short and not very explicit.[15] However, Einstein refers to Hubble's observations throughout the paper and there is little doubt that he was aware of Hubble's value of 500 km/s/Mpc for the rate of expansion.[16]

### 4.1. Einstein's error explained

What is the source of Einstein's error? Luckily, the blackboard casts light on this puzzle. We recall that, in the section of the Rhodes lecture where Einstein turns to astronomical observation to obtain estimates of cosmic parameters, he explicitly wrote $D^2 \sim 10^{-53}$, the fourth relation on the blackboard. This is an important clue. Substituting a value of $H = 500$ km/s/Mpc into the expression $D = H/c$ we obtain $D = 5.4 \times 10^{-28}$ cm$^{-1}$ or $D^2 = 3 \times 10^{-55}$ cm$^{-2}$. Thus, it appears that Einstein erred by a factor of ten in converting the Hubble constant to cgs units.

This impression is confirmed by considering the second equation on the board. Substituting $H = 500$ km/s/Mpc directly into the relation $P \sim 1/D$, we obtain $P = 1.85 \times 10^{27}$ cm or $2 \times 10^9$ light years for the present radius of the Universe, an estimate that is an order of magnitude larger than that on the board (or in the 1931 paper).

Finally, substituting $H = 500$ km/s/Mpc into the relation $D^2 \sim \kappa\rho$, we obtain $\rho \sim 1.6 \times 10^{-28}$ g/cm$^3$ for the density of matter, an estimate that is two orders of magnitude smaller than that on the board (or in the 1931 paper). It is worth noting that our estimates are in agreement with a later review of the 1931 model by Einstein, although he never commented on the correction.[17]

### 4.2. On the timeline of expansion

Einstein's estimate on the blackboard of $10^{10}$ ($10^{11}$) years for the time of cosmic expansion, $t$, is also of interest. Estimates of the time of expansion were a significant problem for early models of the expanding Universe; if interpreted as the age of the Universe, they were often in conflict with estimates of the age of the Earth and of the stars.[18] An estimate of $10^{10}$ years is also given in the 1931 paper, but Einstein does not give an indication of how he arrives at the figure.

It was realized at the time that the inverse of the Hubble constant could give a rough value for this parameter;[19] however, we note that the blackboard line $D^2 = 10^{-53}$ cm$^{-2}$ implies a timeline of expansion of $t \sim 6.2 \times 10^{15}$ s or $t \sim 2 \times 10^8$ years. I have suggested elsewhere that Einstein may have taken his estimate of the time of expansion from Friedmann.[20] In any event, Einstein's value was still problematic in comparison with estimates of the age of the stars, a conflict he attributed in the 1931 paper to the simplifying assumptions of the model: 'One can seek to escape this difficulty by noting that the inhomogeneity of the distribution of stellar material makes our approximate treatment illusory,' he wrote.[21]

### 5. The aftermath

In the aftermath of his visit to Oxford, Einstein received several requests to write up his Rhodes lectures as a short book to be published by Oxford University Press.[22] However, he showed little enthusiasm for the project. Indeed, he is reputed to have later said that he 'had since discovered that everything he had put forward in the lectures was untrue'.[23]

This statement may refer to the second Rhodes lecture, as Einstein proposed an alternative model of the cosmos a year later. On the other hand, it could also refer to the third lecture, as Einstein's approach to unified field theory was in flux at this time. Whatever the reason, Einstein's three Rhodes lectures were never formally published.

However, we suggest that the format of the three lectures almost certainly provided the inspiration for a little-known book on relativity by Einstein published in France in 1933; this book comprised a set of three review articles on general relativity, cosmology, and unified field theory, translated into French by his friend the Romanian mathematician Maurice Solovine (1875–1958).[24] Indeed, in a letter accompanying his manuscript to Solovine, Einstein was careful to reserve the rights for an English publication:

> I managed to tie the thing together only after putting myself to a great deal of trouble and going through much reshuffling and some real work. But now it is crystal clear. I hope you will like it. However, I reserve the right to incorporate it later into an English publication that I have been promising for two years … please return the manuscript after you have finished the translation.[25]

It seems reasonable to assume that the 'English publication that I have been promising for two years' is a reference to the Oxford University Press project. If so, it is a great pity that Einstein did not follow through on his promise, as the French book was not widely distributed. No doubt Einstein had other concerns at the time; just three short months after his letter to Solovine, political events in Germany forced him into a hasty and permanent exile.[26]

### 6. Conclusions

Einstein's Oxford blackboard is well-known as a museum artefact. However, the science underpinning the writing on the board deserves to be better known as it constitutes a visual representation of one of the first models of the expanding Universe. In addition, one line on the blackboard casts useful light on a puzzling systematic error in Einstein's use of astronomical data to derive cosmological parameters from his model: it seems to be due to an order-of-magnitude error in converting the Hubble constant from megaparsecs to cgs units.

We note finally that it has recently been suggested that the writing on the Oxford blackboard might not in fact be Einstein's.[27] I find this hypothesis unlikely as the writing appears to be similar to that of Einstein's handwritten manuscripts and on other blackboards used by him.[28] In addition, none of the contemporaneous reports of Einstein's Rhodes lectures mention an assistant.[29] Finally, three days after the second lecture, Robert Gunther, an Oxford historian of science who founded the History of Science Museum at the university, formally thanked the secretary of the Rhodes trustees for 'your present of two blackboards used by Professor Einstein in his lecture'.[30]

### Acknowledgements
The author thanks Dr Peter Ells and Dr Robert Fox of the History of Science Museum at Oxford for helpful corrections and suggestions. I also thank Andrew Robinson for many useful discussions.

### References and notes

1  See http://www.mhs.ox.ac.uk/object/inv/44725 and associated links.

2  Fox, Robert, 'Einstein in Oxford', *Notes and Records of the Royal Society of London*, 72 (2018), 293–318; Robinson, Andrew, 'Einstein in Oxford', *Physics World*, 32 (2019), 50–55.

3  See https://www.bbc.com/news/av/uk-england-oxfordshire-41096926/the-blackboard-albert-einstein-left-in-oxford-in-the-1930s

4  O'Raifeartaigh, Cormac, and McCann, Brendan, 'Einstein's Cosmic Model of 1931 Revisited; An Analysis and Translation of a Forgotten Model of the Universe', *European Physical Journal H*, 39 (2014), 63–85; see also Kragh, Helge, 'Cyclic Models of the Relativistic Universe: the Early History', in *Einstein Studies Vol. 14; Beyond Einstein: Perspectives on Geometry, Gravitation and Cosmology in the Twentieth Century*, ed. Rowe, David E., Sauer, Tilman, and Walter, Scott A. (Birkhäuser, 2018), 183–204.

5  Albert Einstein, 'Zum kosmologischen Problem der allgemeinen Relativitätstheorie', *Sitzungsberichte der Königlich Preussischen Akademie der Wissenschaften* (1931), 235–7.

6  Fox, op. cit. (ref. 2); Robinson, Andrew, *Einstein on the Run; How Britain Saved the World's Greatest Scientist* (Yale University Press, 2019), 145–63.

7   A summary in English of the content of the lectures was given to members of the audience as a pamphlet. The text of this pamphlet can be viewed at http://www.mhs.ox.ac.uk/collections/imu-search-page/narratives/?irn=9738&index=1

8   We know from his diary that Einstein was not pleased by the fuss. For more on the incident see Robinson, op. cit. (ref. 6), 160–1. See also Eisinger, Josef, *Einstein On the Road* (Prometheus Books, 2011), p. 128.

9   Einstein, op. cit. (ref. 5).

10  Einstein added the so-called cosmological constant term to the field equations of relativity in 1917 in order describe a universe that was assumed to be static. With the emergence of the first evidence of cosmic expansion, he saw the term as redundant. For details see O'Raifeartaigh and McCann, op. cit. (ref. 4) or Nussbaumer, Harry, 'Einstein's Conversion to the Expanding Universe', *European Physical Journal H*, 39 (2014), 37–62.

11  Einstein, Albert, and de Sitter, Willem, 'On the Relation Between the Expansion and the Mean Density of the Universe', *Proceedings of the National Academy of Sciences*, 18 (1932), 213–4.

12  Einstein, op. cit. (ref. 5).

13  Although not stated on the board, the units for this quantity are $cm^{-2}$ as Einstein used cgs units for all such calculations (ref. 3).

14  Einstein and de Sitter, op. cit. (ref. 11).

15  Einstein, op. cit. (ref. 5).

16  Nussbaumer, Harry, and Bieri, Lydia, *Discovering the Expanding Universe* (Cambridge University Press, 2009), 145–6.

17  Einstein, Albert, 'On the cosmologic problem', Appendix I to *The Meaning of Relativity* (Princeton University Press, 1945, 3rd ed.), 130–45.

18  Kragh, Helge, *Cosmology and Controversy* (Princeton University Press, 1996), 73–76; Nussbaumer and Bieri, op. cit. (ref. 16), 153–6.

19  Ibid.

20  O'Raifeartaigh and McCann, op. cit. (ref. 4).

21  Einstein, op. cit. (ref. 5).

22  See Fox, op. cit. (ref. 2); Robinson, op. cit. (ref. 6), 159–60.

23  Ibid.

24  Einstein, Albert, *La Théorie de la Relativité* (Hermann et Cie, 1933, trans. M. Solovine).

25  Einstein, Albert, letter to Maurice Solovine 1932 September 29, in *Albert Einstein: Letters to Solovine 1906–1955* (Citadel Press, 1993), p. 74.

26  Robinson, op. cit. (ref. 6), 192–3.

27  Graham Farmelo, private communication, 2020 May.

28  See for example Document [1-115] on the *Albert Einstein Online Archive* http://aefind.huji.ac.il/vufind1/Record/EAR000034123

29  Brief reports on the lectures appeared in *The Times* on 1931 May 11, 18, and 25 and in *Nature* on 1930 May 16, 23, and 30. See Fox op. cit. (ref. 2) for further details.

30  Robert Gunther, letter to the Rhodes trustees, 1931 May 19. The second blackboard was unfortunately wiped clean. For details, see Robinson, op. cit. (ref. 6), 162–3.

**The author**

Cormac O'Raifeartaigh (Cormac O'Rafferty) is an Irish physicist based at Waterford Institute of Technology in Ireland. A solid-state physicist by training, he is best known for several contributions to the study of the history and philosophy of 20th-century science, including the discovery that Albert Einstein once attempted a steady-state model of the expanding universe, many years before Fred Hoyle. He was elected a Fellow of the Royal Astronomical Society in 2014 and a Fellow of the Institute of Physics in 2016. He is also an Adjunct Research Fellow at the Dublin Institute for Advanced Studies and a Visiting Associate Professor at the School of Physics at University College Dublin. He blogs at https://antimatter.ie